\documentclass[12pt]{iopart}
\usepackage{iopams}  
\usepackage{bm}
\usepackage{graphicx}
\bibstyle{approve.bib}

\newcommand{\be}{\begin{equation}}
\newcommand{\ee}{\end{equation}}
\newcommand{\bea}{\begin{eqnarray}}
\newcommand{\eea}{\end{eqnarray}}

\newcommand{\bom}{\mathbf{\Omega}}

\newcommand{\ep}{\varepsilon}

\newcommand{\te}{\tilde{\varepsilon}}

\newcommand{\tchi}{\tilde{\chi}}
\newcommand{\tG}{\tilde{G}}

\newcommand{\tP}{\tilde{P}}
\newcommand{\tE}{\tilde{E}}

\begin{document}

\title[Nonlocal and nonlinear electrostatics of a dipolar Coulomb fluid]{Nonlocal and nonlinear electrostatics of a dipolar Coulomb fluid}

\author{Sahin Buyukdagli and Ralf Blossey}

\address{Interdisciplinary Research Institute, Universit\'e des Sciences et des Technologies
de Lille (USTL), USR CNRS 3078, 50 Avenue Halley, 59568 Villeneuve d'Ascq, France}

\ead{ralf.blossey@iri.univ-lille1.fr}
\begin{abstract}
We study a model Coulomb fluid consisting of dipolar solvent molecules of finite extent generalizing the point-like
Dipolar Poisson-Boltzmann model (DPB) previously introduced by Coalson and Duncan (J. Phys. Chem {\bf 100}, 2612 (1996))
and Abrashkin {\it et al.} (Phys. Rev. Lett. {\bf 99}, 077801 (2007)). We
formulate a nonlocal Poisson-Boltzmann equation (NLPB)  and study both linear and nonlinear dielectric response
in this model for the case of a single plane geometry. Our results shed light on the relevance of nonlocal vs nonlinear effects
in continuum models of material electrostatics.
\end{abstract}


\date{\today}                                           
\maketitle

Electrostatic properties of soft matter or biological molecules and cellular structures are essential factors for their function. In the last thirty
years there have been numerous applications of continuum electrostatics to such systems \cite{bruinsma02,nel09}. The vast majority of them have,
in one way or another, relied on `homogeneous' continuum electrostatics which assumes that the polar solvent in which a charged structure is
embedded can be characterized by a dielectric constant,  $\varepsilon \approx 80 $ in the case of water. The corresponding Poisson or Poisson-Boltzmann
equations are then solved within mean-field theory. This approach obviously fails when molecular length scales become truly relevant, as this is, e.g.,  the
case for the solvation of proteins. Two major avenues have been developed in recent years to improve upon this macroscopic, structureless treatment.

i) {\it Nonlocal electrostatics.} Nonlocal electrostatics refers to an originally phenomenological approach pioneered by the Russian school dating back to
the 1980's in which the medium is assumed to be described by a spatially-dependent dielectric function, $\varepsilon = \varepsilon({\bf r}, {\bf r'})$. This approach
allows to include the complex spatial dependence of the dielectric behavior of water, although for calculations often approximate models
have been employed \cite{kornyshev85,bopp96,kornyshev97-1,kornyshev97-2}. Recently, one of us (SB) has developed a microscopic theory of
nonlocal electrostatics of Coulomb fluids~\cite{buyukdagli13}.

ii) {\it Nonlinear structured electrostatics.} Nonlinear electrostatic models have been developed which
include details of the water structure due to the dipolar nature of the solvent which are assumed as point-like.
These theories lead to Poisson-Boltzmann  equations with more complex nonlinearities than in the homogeneous case
\cite{coalson96,abrashkin07,azuara08}.

In addition, the mean-field Poisson-Boltzmann approach is know to fail for highly charged systems, as is often the case in biologically relevant settings. For this case,
the so-called `strong-coupling limit' (SC), and different approximate schemes have been developed, like the one-loop approximation and  variational methods
that go beyond the mean-field solutions (for a recent review, see \cite{naji13}).
These approaches have so far also relied on a homogeneous, i.e. structureless, medium as their starting point.
It is, however, to be expected that also beyond mean-field theory, fluid structure will become of relevance. Therefore it is important to have a clear idea of
the relationship between the approaches i) and ii) already at the mean-field level.
For this aim we study a simplified version of the model treated in \cite{buyukdagli13} in which only the dipolar degrees of freedom of the solvent molecules are considered.
This model is thus the simplest nonlocal extension of the Dipolar Poisson-Boltzmann model which has seen some discussion in the literature.
We derive the nonlocal  Poisson-Boltzmann equation (NLPB) for the solvent model and determine its dielectric response first in the linear regime and then in the nonlinear regime that had not been considered in our previous work~\cite{buyukdagli13}.

We begin by recalling the basics of the equations of nonlocal electrostatics within the macroscopic phenomenological approach and explain the link to the
microscopic theory. The natural starting point  is Gauss' law
$ \nabla \cdot {\bf D} = \varrho $ where the dielectric displacement field is given, as usually, by ${\bf D} =
{\bf E} \,+\, {\bf P}$. Hence  $ \nabla \cdot {\bf E} + \nabla \cdot {\bf P} = \varrho\, $
where ${\bf E} $ is the electric field and ${\bf P}$ the polarization field. Since $ {\bf E} = - \nabla \phi $ one has the standard relation between the electrostatic
potential $\phi$ and the polarization field
${\bf P}$,
\begin{equation}  \label{one}
-\Delta \phi + \nabla \cdot {\bf P} = \varrho\, .
\end{equation}
This is nothing but the Poisson-Boltzmann equation if $\varrho$ contains both fixed (surface-) and mobile (ionic) charges. The $\nabla \cdot {\bf P} $-term is the source of the
polarization charges.

`Nonlocality' comes into play in the relation between ${\bf P} $ and ${\bf E}$ which specifies the material (i.e., solvent) properties. Generally this relation is written as $ {\bf P} = \chi {\bf E} $
where $\chi$ is the susceptibility. In a local theory, $\chi$ is a constant, while in a nonlocal theory it is in an integral operator of the form
\begin{equation}   \label{nl}
{\bf P}({\bf r}) =  [\chi {\bf E}]({\bf r})  \equiv \int d\bf{r}'\; \chi({\bf r},{\bf r'}) {\bf E}({\bf r'})\, .
\end{equation}
In the literature, two approaches have been proposed to render the theory local. One of them \cite{hildebrandt04}, considers the integral kernel
$\chi({\bf r},{\bf r'})$ as a Green function to a differential operator ${\cal L}$ via $ {\cal L} \cdot {\cal \chi} = \delta $. Eq.(\ref{nl}) then is turned into a local
relation $ {\cal L} {\bf P} = - \nabla \phi\, $, which holds in the linear case and obviously is the inverse of equation (\ref{nl}).
In Ref. \cite{maggs06} it was shown that a more general form of the equation can be derived from a local polarization functional $U_P[{\bf P}]$ of
Ginzburg-Landau type such that
\begin{equation} \label{two}
-\nabla \phi = \frac{\delta U_P[{\bf P}]} {\delta{\bf P}}\, .
\end{equation}

In the case of a microscopic theory, eq.~(\ref{one}) still holds, but there is no equivalent of eq.~(\ref{two}), and rather a generalization of eq.~(\ref{nl})
arises. This is easily illustrated in the case of a  dipolar solvent.
For this we introduce a Coulomb fluid model composed of a symmetric electrolyte of two ionic species with valencies $q_\pm=\pm q$ and $q>0$, immersed in a polar solvent. The solvent consists of linear dipole molecules of finite size, each composed of two elementary charges of opposite sign $+Q$ and $-Q$ separated by a fixed distance $a$. One can derive the partition function as  $ Z_G=\int \mathcal{D}\phi\;e^{-H[\phi]} $
with the Hamiltonian functional \cite{buyukdagli13}
\bea\label{HamFunc}
H[\phi]&=&\int d{\bf r}\left[\frac{\left[\nabla\phi(\bf{r})\right]^2}{8\pi\ell_B}- {\it i}\sigma(\bf{r})\phi(\bf{r})\right]\\
&&-\Lambda_s\int\frac{d{\bf r}d\bom}{4\pi}e^{iQ\left[\phi({\bf r})-\phi({\bf r}+{\bf a})\right]} -\int d {\bf r}\left\{\Lambda_+e^{{\it i}q\phi({\bf r})}+ \Lambda_-e^{-{\it i}q\phi(\bf{r})}\right\},\nonumber
\eea
where the ionic and solvent fugacities are scaled as $\Lambda_i=e^{\mu_i}/\lambda_{Ti}^3$ and $\Lambda_s=e^{\mu_s}/\lambda_{Td}^3$
with the corresponding chemical potentials and thermal wavelengths. Furthermore, in Eq.~(\ref{HamFunc}), the temperature is included in the Bjerrum length in air, given by  
$\ell_B=e^2/(4\pi\ep_{air} k_BT)\simeq 54.6$ nm, and $\sigma({\bf r})$ is the
fixed charge distribution. We note that if we expand the exponential of the dipolar term for ${\bf a} \rightarrow 0$, the model reduces to the dipolar Poisson-Boltzmann model (DPB).

By passing from the complex to the real electrostatic potential via the transformation $\phi({\bf r})\to i\phi({\bf r})$, the MF-level saddle point equation $ \delta H[\phi]/\delta \phi(\bf{r})=0 $ in the case of a permeable charge
distribution of planar geometry $\sigma({\bf r})=-\sigma_s\delta(z)$ takes the form of a nonlocal Poisson-Boltzmann (NLPB) equation,
\bea\label{nlpb}
&&\Delta\phi(z)+4\pi\ell_B\sigma(z)-8\pi\ell_B\rho_i^b q\sinh\left[q\phi(z)\right]\\
&&+8\pi\ell_BQ\rho_s^b\int_{-a}^a\frac{da_z}{2a}\sinh\left[Q\phi(z+a_z)-Q\phi(z)\right]=0,\nonumber
\eea
where we used the MF relations $\Lambda_i=\rho_i^b$ and $\Lambda_s=\rho_s^b$~\cite{buyukdagli13} for the ionic and dipolar bulk densities, respectively. The comparison with the general structure of the Poisson-Boltzmann equation shows that we have for the polarization density
\begin{equation}\label{polc}
\frac{\partial P(z)}{\partial z} = n_{sc}(z)
= 2Q\rho_{sb}\int_{-a}^a\frac{da_z}{2a}\sinh\left[Q\phi(z+a_z)-Q\phi(z)\right],
\end{equation}
i.e., an integral relationship between ${\bf P}$ and the potential $\phi$.

In the regime of weak surface charges where one has $\phi(z)<1$, linearizing the NLPB equation~(\ref{nlpb}) we get
\begin{equation}\label{nlpb3x}
\Delta\phi_0(z)-\kappa_i^2\phi_{\bf 0}(z)+\kappa_s^2\int_{-a}^a\frac{da_z}{2a}\left[\phi_0(z+a_z)-\phi_0(z)\right] =-4\pi\ell_B\sigma(z),
\end{equation}
where the index `0' means that we dropped the non-linear corrections. Furthermore, we introduced the ionic and solvent screening parameters in the air medium as $\kappa_i^2 = 8\pi\ell_B\rho_i^bq_i^2$ and
 $\kappa_s^2 = 8\pi\ell_B\rho_s^bQ^2$, respectively. Solving Eq.~(\ref{nlpb3x}) in Fourier space, one gets in the regime $0\leq \kappa_iz\ll1$ the electric field $E(z)=\partial_z\phi(z)$ in the form
\be\label{net}
E(z)=\frac{2\pi\ell_B\sigma_s}{\ep_{eff}(z)},
\ee
which defines the local effective dielectric permittivity
\be\label{peref}
\ep_{eff}(z)=\frac{\pi}{2}\left/\int_0^\infty\frac{\mathrm{d}k}{k}\frac{\sin(kz)}{\te(k)}\right.
\ee
with the Fourier-transformed permittivity $\te(k)=1+4\pi\ell_B\tchi_0(k)$ and the susceptibility function
\be\label{eq4}
\tchi_0(k)=\frac{\kappa_s^2}{4\pi\ell_Bk^2}\left[1-\frac{\sin(ka)}{ka}\right],
\ee
Eq.~(\ref{peref}) clarifies the concept of a distance-dependent effective permittivity which is often used phenomenologically in the literature, see, e.g. \cite{mallik02}.
To leading order in the solvent density $O\left((\kappa_sa)^2\right)$, the effective permittivity follows from Eq.~(\ref{peref}) in the simple form
\be\label{efasym}
\ep_{eff}(z)=1+\frac{\left(\kappa_sa\right)^2}{6}\left\{1-\left(1-\frac{z}{a}\right)^3\theta(a-z)\right\},
\ee
where $\theta(z)$ stands for the Heaviside function. Thus, for dilute solvents and in the linear response regime, the effective permittivity decreases from the bulk permittivity $\ep_w=1+\left(\kappa_sa\right)^2/6$ to the air permittivity at the interface where the polarization field vanishes. It should be noted that the same dielectric reduction effect at charged interfaces has been constantly observed in molecular dynamics simulations with explicit solvent (see e.g. Ref.~\cite{hansim}) and in theoretical formulations based on integral equations~\cite{blum}. With the use of diffuse permittivity functions such as the Inkson dielectric model~\cite{ink}, this peculiarity has been also artificially introduced into phenomenological formulations of non-local electrostatic interactions. The present formalism clearly indicates that the dielectric reduction effect is a direct consequence of the finite solvent molecular size responsible for the  non-local response of the fluid to the charged interface.

We note that the linear response relations~(\ref{nlpb3x})-(\ref{efasym}) have been previously introduced in our recent article~\cite{buyukdagli13}. We now continue on to the nonlinear case that had not been treated in our previous work. To obtain the non-linear response relation, we first have to derive the non-linear susceptibility function, which we will do perturbatively. To this end, we first insert into the non-linear MF free energy~(\ref{HamFunc}) the Ansatz $\phi(z)=\int_{-\infty}^{+\infty}\mathrm{d}z'G(z-z')\sigma(z')$, and expand the equation to second order in the surface charge. Further, we write the Green function as the expansion 
\be\label{per1}
G(z)=G_0(z)+\lambda G_1(z),
\ee
where the perturbative parameter $\lambda$ will allow to compute the correction to the linear response solution, and the Green's  function $G_0(z)=\int\mathrm{d}k\;\tG_0(k)/(2\pi)$ solves the linear NLPB equation~(\ref{nlpb3x}) in the form $\phi_0(z)=\int_{-\infty}^{+\infty}\mathrm{d}z'G_0(z-z')\sigma(z')$, with
\be\label{grlin}
\tG_0^{-1}(k)=\frac{\kappa_i^2+k^2}{4\pi\ell_B}+k^2\tchi_0(k),
\ee
Then, one finds that the correction to the Green's function satisfies the differential equation
\bea\label{eqexp2}
&&\partial_z^2G_1(z)-\kappa_i^2G_1(z)-\kappa_s^2\int_{-a}^a\frac{da_z}{2a}\left[G_1(z)-G_1(z+a_z)\right]\nonumber\\
&&=\frac{\sigma_s^2}{6}\left\{q^2\kappa_i^2G_0^3(z)+\kappa_s^2Q^2\int_{-a}^{+a}\frac{\mathrm{d}a_z}{2a}\left[G_0(z)-G_0(z+a_z)\right]^3\right\}\nonumber\\
\eea
Solving Eq.~(\ref{eqexp2}) in Fourier space, the non-linear correction follows as
\be\label{eqexp4}
\tG_1(k)=-\frac{\sigma_s^2}{24\pi\ell_B}\tG_0(k)\left\{q^2\kappa_i^2F(k)+Q^2\kappa_s^2T(k)\right\},
\ee
where we defined the functions
\bea
F(k)&=&\int_{-\infty}^{+\infty}\frac{\mathrm{d}k_1\mathrm{d}k_2}{4\pi^2}\tG_0(k_1)\tG_0(k_2)\tG_0(k-k_1-k_2\nonumber)\\
&&\\
T(k)&=&\int_{-\infty}^{+\infty}\frac{\mathrm{d}k_1\mathrm{d}k_2}{4\pi^2}\tG_0(k_1)\tG_0(k_2)\tG_0(k-k_1-k_2)\nonumber\\
&&\hspace{3.1cm}\times R(k_1,k_2,k-k_1-k_2),\nonumber\\
\eea
with the structure factor
\bea
R(k_1,k_2,k_3)&=&1-\sum_{i=1}^3\frac{\sin(k_ia)}{k_ia}+\frac{\sin\left[(k_1+k_2)a\right]}{(k_1+k_2)a}\nonumber\\
&&+\frac{\sin\left[(k_1+k_3)a\right]}{(k_1+k_3)a}+\frac{\sin\left[(k_2+k_3)a\right]}{(k_2+k_3)a}\nonumber\\
&&-\frac{\sin\left[(k_1+k_2+k_3)a\right]}{(k_1+k_2+k_3)a}.
\eea
\begin{figure}
\begin{center}
\resizebox{0.65\textwidth}{!}{%
\includegraphics{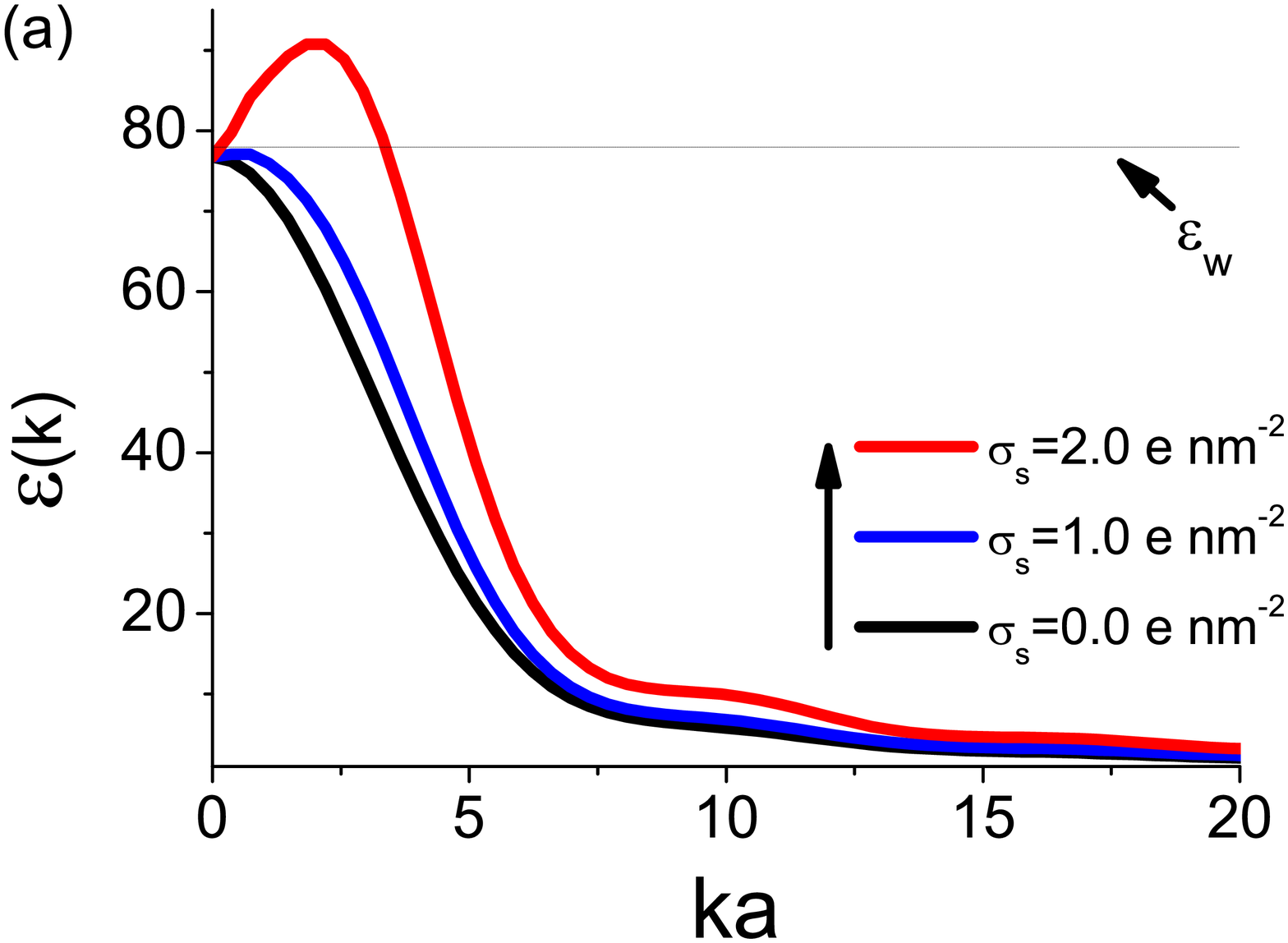}
}
\\
\resizebox{0.65\textwidth}{!}{%
\includegraphics{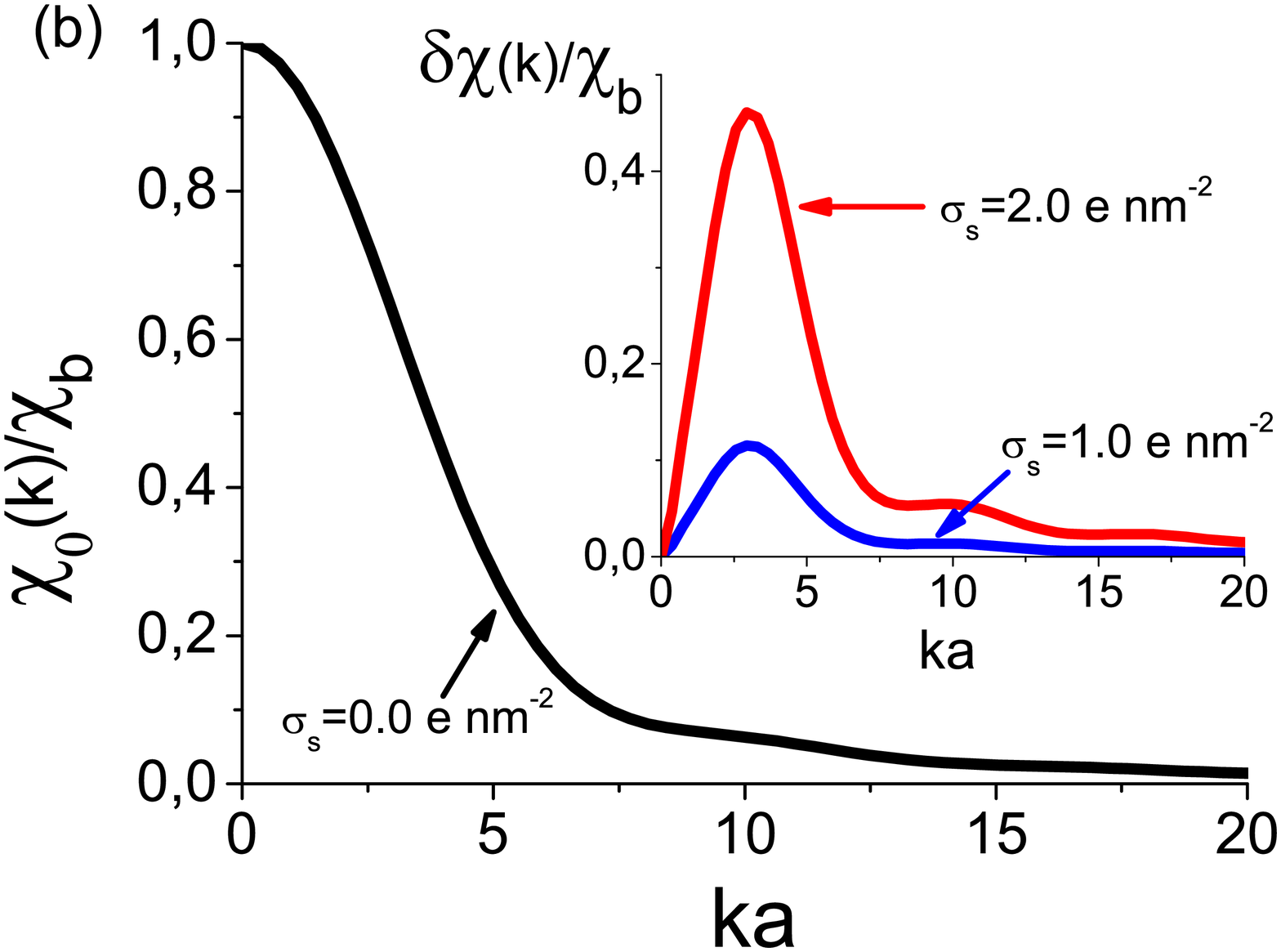}
}
\\
\resizebox{0.65\textwidth}{!}{%
\includegraphics{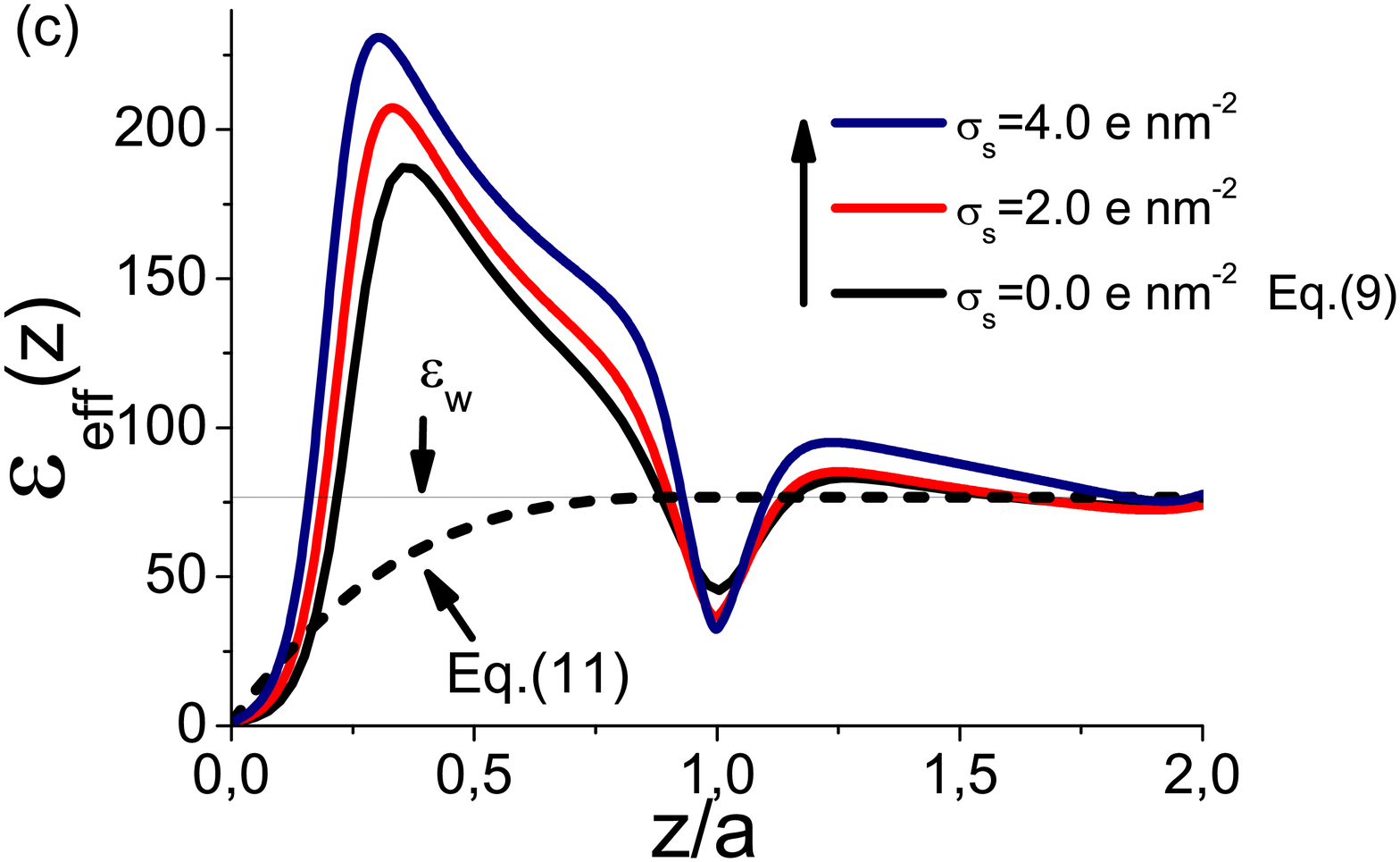}
}
\end{center}
\caption{(a) Surface charge dependence of the dielectric permittivity function of Eq.~(\ref{disig}) and (b) the linear (main plot) and non-linear response part (inset) of the susceptibility function in Fourier space, normalized by the susceptibility of the dielectric continuum electrostatics $\chi_b=\rho_{sb}Q^2a^2/3$. (c) Dielectric permittivity profile in real space. Model parameters are $a=1$ {\AA}, $Q=1$, $\rho^b_s=55$ M, $\rho_i^b=0$ M, which results in the bulk permittivity $\ep_w=77$ (horizontal line in (a) and (c)).}
\label{fig2}       
\end{figure}
At the perturbative order $O(\lambda)$ the Fourier transform of the kernel associated with the Green's function~(\ref{per1}) can be written as
$\tG^{-1}(k)=\tG_0^{-1}(k)-\lambda\tG_0^{-2}(k)\tG_1(k)$. Taking into account Eq.~(\ref{grlin}) and the non-linear contribution in Eq.~(\ref{eqexp4}), one finds that the Fourier transformed kernel can be expressed in the form
\be\label{eqexp6}
\tG^{-1}(k)=\frac{\kappa_i^2+\lambda\delta\kappa_i^2(k)+k^2}{4\pi\ell_B}+k^2\left[\tchi_0(k)+\lambda\delta\tchi(k)\right],
\ee
with the screening and the susceptibility functions associated with the non-linearities,
\bea\label{eqexp7}
\delta\kappa_i^2(k)&=&\frac{q^2\sigma_s^2\kappa_i^2}{6}\frac{F(k)}{\tG_0(k)}\\
\label{eqexp72}
\delta\tchi(k)&=&\frac{Q^2\sigma_s^2\kappa_s^2}{24\pi\ell_B}\frac{T(k)}{k^2\tG_0(k)}.
\eea

In order to derive the non-linear dielectric response relation, we finally reconsider Eq.~(\ref{polc}). From now on, we will set $\lambda=1$. Injecting into the rhs of this equation the potential in the
form $\phi(z)=-\sigma_s\left[G_0(z)+G_1(z)\right]$, expanding up to the cubic order in the surface charge, and Fourier-transforming the expansion, one gets for the polarization field
\be\label{eqexp9}
\tP(k)=\tchi_0(k)\left[\tE_0(k)+\tE_1(k)\right]+\delta\tchi\left[\tE_0;k\right]\tE_0(k),
\ee
where we introduced the electric field in the linear response regime $\tE_0(k)=-ik\sigma_s\tG_0(k)$ and the non-linear correction $\tE_1(k)=-ik\sigma_s\tG_1(k)$, and we recast the non-linear part of the susceptibility function~(\ref{eqexp72}) into a more intuitive form without explicit dependence on the surface charge,
\bea\label{eqexp10}
\delta\tchi\left[\tE_0;k\right]&=&-\frac{\rho_s^bQ^4}{3k^4\tG_0(k)}\int_{-\infty}^{+\infty}\frac{\mathrm{d}k_1\mathrm{d}k_2}{4\pi^2}\tG_0(k-k_1-k_2)\nonumber\\
&&\hspace{2.6cm}\times R(k_1,k_2,k-k_1-k_2)\nonumber\\
&&\hspace{2.6cm}\times\tE_0(k_1)\tE_0(k_2).
\eea
One sees in Eq.~(\ref{eqexp10}) that at the leading perturbative order the non-linear response translates into a quadratic dependence of the susceptibility function on the electric field. This dependence stems from the fact that a change in
the sign of the electric field is expected in Eq.~(\ref{eqexp9}) to result in the change of the sign of the reaction field.

We will now scrutinize non-linear effects on the dielectric permittivity. Making use of Eq.~(\ref{eqexp6}), one can express the Fourier transformed dielectric permittivity function in the non-linear response regime as
\be\label{disig}
\te(k)=1+4\pi\ell_B\left[\tchi_0(k)+\delta\tchi(k)\right]
\ee
The relation~(\ref{disig}) with Eq.~(\ref{eqexp72}) shows that at the leading order, the dielectric permittivity has a quadratic dependence on the surface charge. We illustrate in Fig.(\ref{fig2})(a) the charge dependence of the permittivity function. First of all, it is seen that the modification of the permittivity by the surface charge becomes significant in the regime $\sigma_s\geq 1$ $\mathrm{e\;nm}^{-2}$, the boundary value corresponding to the characteristic charge of the DNA molecule known to be located in the intermediate electrostatic coupling regime. This means that the non-linear response behavior comes into play in the surface charge regime where correlation effects absent in the present formulation should be included. Then, one notes that the permittivity increases with the surface charge, i.e. the dielectric saturation effect is not observed in our model. Finally, one notices that the surface charge makes no contribution to the permittivity function in the infrared ($k\to 0$) and ultraviolet regime ($k\to\infty$) corresponding to the bulk region and the close vicinity of the interface, respectively.

To explain the dependence of the Fourier transformed dielectric permittivity on the surface charge in the non-linear regime, we compare in Fig.\ref{fig2}(b) the linear response susceptibility (main plot) and the non-linear response contribution (inset). In agreement with the permittivity curves in Fig.\ref{fig2}(a), the non-linear response contribution is seen to increase the linear susceptibility, and this contribution is present only at intermediate wavelengths. The latter point becomes clear if one notes that according to Eq.~(\ref{efasym}), the close vicinity of the interface corresponds to a dielectric void where the polarization field vanishes. In the opposite limit $k\to0$ corresponding to large separations from the interface, non-linear contributions are expected to die out. As a result, non-linear effects do not modify the dielectric response in both regimes. In particular, one sees that the amplification of the susceptibility at high surface charges results in a maximum at intermediate wave vectors, which in turn leads to a peak in the Fourier-transformed permititvity profile at the surface charge $\sigma_s=2.0$ $\mathrm{e\;nm}^{-2}$. Indeed, the physical mechanism behind the rise in the susceptibility with the surface charge can be easily understood by noting that using Eq.~(\ref{polc}) with Eqs.~(\ref{eqexp4}) and~(\ref{eqexp9}), one can relate in Fourier basis the surface charge dependence of the susceptibility function to the non-linear response part of the solvent charge density as
\be\label{nlnsc}
\tilde{n}_{sc}(k)=\tilde{n}^{(0)}_{sc}(k)\left[1+\frac{k^2\tG_0(k)}{4\pi\ell_B}\frac{\delta\tchi(k)}{\tchi_0(k)}\right],
\ee
where the linear response part of the solvent charge density reads $\tilde{n}^{(0)}_{sc}(k)=k^2\tchi_0(k)G_0(k)\sigma_s$. The relation~(\ref{nlnsc}) indicates that non-linearities amplify the solvent charge density according to a cubic dependence on the surface charge. This in turn increases the amplitude of the polarization field (see Eq.~(\ref{polc})) and the importance of the dielectric screening effect  resulting in a larger dielectric permittivity in Fig.(\ref{fig2})(a).

We finally show in Fig.~\ref{fig2}(c) the dependence of the effective dielectric permittivity on the distance from the charged interface. The comparison of the dashed and solid black curves corresponding to the vanishing surface charge limit $\sigma_s\to0$ shows that non-local effects associated with the departure from the dilute solvent regime translate into fluctuations of the local permittivity function around the bulk permittivity. Then, to illustrate the non-linear dielectric response behavior of the solvent in real space, we reported in Fig.~\ref{fig2}(c) the dielectric permittivity profile for finite surface charges. The latter was obtained via Eq.~(\ref{net}) from the numerical solution of the non-linear NLBP equation~(\ref{nlpb}) in the limit of vanishing salt density $\rho_{ib}\to0$ (see ~\ref{ap1} for the details of the numerical relaxation algorithm). In qualitative agreement with the Fourier transformed permittivity profiles in Fig.~\ref{fig2}(a), the increase of the surface charge resulting in a deviation from the linear response regime is seen to increase the amplitude of the dielectric permittivity in real space. One also notes that interestingly, the periodicity of the effective permittivity profile is not modified by the surface charge.

To conclude we have clarified the relationship between phenomenological and microscopic approaches to nonlocal electrostatics in structured Coulomb fluids. For the case of the nonlocal Poisson-Boltzmann equation
for a dipolar solvent, we have reviewed the linear response properties of the liquid and also determined its nonlinear dielectric response behaviour that we had not considered in Ref.~\cite{buyukdagli13}. Nonlinear contributions become significant in regimes in which fluctuation effects can usually not be neglected. Thus, nonlocal theories
of material electrostatics can be seen as being essentially relevant to linear regimes in cases where structural effects matter. It will be interesting in the future to understand the role of nonlocal electrostatics beyond
mean-field theory. 

It is useful to indicate the limitations and possible extensions of our model. We emphasize that MD simulations with confined solvents exhibit the dielectric anisotropy effect characterised by a vector form of the effective permittivities~\cite{hansim}. The scalar form of the permittivity in the present mean-field formulation clearly results from the absence of solvent-membrane correlations expected to break the spherical anisotropy. Thus, the inclusion of solvent correlations into the present approach should enable us to cover the anisotropic dielectric response of the liquid. Within our model, we have not found either the dielectric catastrophe effect observed in numerical simulations and phenomenological non-local formulations of electrostatics~\cite{bopp96,bont2}. The proper account for the hydrogen bonding between the solvent molecules may be necessary to include this peculiarity. Moreover, considering the high concentration of water solvent, one should extend in a future work the NLPB approach by including hard-core interactions between solvent molecules. These interactions are expected to result in a wetting of the interface by solvent molecules, which may increase the dielectric permittivity profiles in Figs.~\ref{fig2}(a)-(c). We finally note that in the present work, we neglected the multipolar moments of solvent molecules. Although water is known to possess high multipolar moments, the multipoles were shown in Ref.~\cite{buyukdagli13} to bring a minor contribution to the dielectric permittivity of the liquid at the mean-field level. The accurate consideration of multipolar effects on the dielectric response of the liquid may thus necessitate the inclusion of solvent fluctuations.
\\

{\bf Acknowledgement.} SB gratefully acknowledges support under the ANR blanc grant ``Fluctuations in Structured Coulomb Fluids''.
\\

\smallskip
\appendix
\section{Relaxation algorithm for the solution of the non-linear NLPB equation}
\label{ap1}
In this appendix, we introduce a relaxation algorithm for the solution of the non-linear NLPB equation~(\ref{nlpb}). The solvent is symmetrically partitioned around the planar interface located at $z=0$ and carrying the surface charge $\sigma_s$. In order to obtain the first boundary condition associated with Eq.~(\ref{nlpb}), one has to integrate this equation in the vicinity of the interface. This gives for the surface field
\be\label{boun}
\phi'(0)=2\pi q_i\ell_B\sigma_s,
\ee
which is simply Gauss' law in air. The second boundary condition is a vanishing electrostatic potential in the bulk, that is $\phi(z\to\infty)=0$.

The numerical scheme consists in solving Eq.~(\ref{nlpb})  on a discrete lattice located between $z=0$ and $z=z_{max}$, and composed of $2N+1$ mesh points separated by the distance $\epsilon$.  We first define the potential on the lattice as $\phi_n\equiv\phi(z_n)$, where the index $n$ running from $1$ to $2N+1$ denotes the position on the lattice, with the discrete distance from the interface $z_n=(n-1)\epsilon$. Using the finite difference form of the Laplacian of the potential in Eq.~(\ref{nlpb}), i.e. $\epsilon^2\phi''(z)=-2\phi_n+\phi_{n+1}+\phi_{n-1}$, the NLPB equation can be rearranged in the form
\bea
\label{nlpbdis}
\phi_n&=&\frac{1}{2}\left\{\phi_{n+1}+\phi_{n-1}-r\sinh\left[q_i\phi_n\right]\right.\nonumber\\
&&\left.+s\sum_{j=j_1(n)}^{j_2(n)}\sinh\left[Q\left(\phi_j-\phi_n\right)\right]\right\},
\eea
where we introduced the coefficients $r=\epsilon^2\kappa_i^2$ and $s=\epsilon^3\kappa_s^2/(2a)$, the auxiliary functions $j_1(n)=n-n_a+1$ and $j_2(n)=n+n_a-1$, with the index $n_a$ defined as $z_{n_a}=a$. In the present work, we will need the solution for the salt free solvent, which gives $r=0$ and removes the first hyperbolic sinus function on the r.h.s. of Eq.~(\ref{nlpbdis}). Moreover, we note that the symmetrical partition of the solvent and the boundary condition Eq.~(\ref{boun}) result, respectively, in the following relations that should be coupled with Eq.~(\ref{nlpbdis}),
\bea
&&\phi_{-n}=\phi_n, \\
&&\phi_0=\phi_1-2\pi q_i\ell_B\sigma_s\epsilon.
\eea

The relaxation algorithm consists in iterating Eq.~(\ref{nlpbdis}) by injecting first into the r.h.s. a guess solution. This yields the updated potential profile $\left\{\psi_n\right\}_n$, which is substituted again into the r.h.s. of the equation and the cycle is continued until numerical convergence is achieved. The key point is that for the convergence of the algorithm, the guess potential used at the first iterative level should also satisfy the boundary condition~(\ref{boun}). Because the solution of the local PB equation obeys a different boundary condition, namely $\phi'(0)=2\pi q_i\ell_w\sigma_s$ with $\ell_w=\ell_B/\ep_w$ the Bjerrum length in water, the latter cannot be used as the reference potential. In order to derive the adequate reference potential, we note that in Ref.~\cite{buyukdagli13}, it was shown that  the linear form of Eq.~(\ref{nlpb}) gives for $0\leq z\leq d_1$ an electrostatic field profile obeying an exponential damping as $\phi'(z)=2\pi\ell_B\sigma_se^{-\kappa_sz}$, with the characteristic distance $d_1=\log(\ep_w)/\kappa_s$.  This gives for the potential $\phi(z)=c_1+2\pi\ell_B\sigma_se^{-\kappa_sz}/\kappa_s$ in this region, with $c_1$ a constant. Then, far from the interface where non-local dielectric response effects vanish, we expect the potential to behave as the solution of the simple Poisson equation in a salt free liquid, i.e. $\phi(z)=2\pi q_i\ell_w\sigma_s z$. Imposing the continuity of the electrostatic potential at $z=d_1$, one finally gets for the reference potential the following piecewise form
\bea
\phi(z)&=&\left\{2\pi q_i\ell_w\sigma_s\left(d_1+\kappa_s^{-1}\right)-\frac{2\pi q_i\ell_B\sigma_s}{\kappa_s}e^{-\kappa_s |z|}\right\}\theta(d_1-|z|)\nonumber\\
&&+2\pi q_i\ell_w\sigma_s z\;\theta(|z|-d_1).
\eea
\\

\end{document}